\documentclass[pra,twocolumn,superscriptaddress]{revtex4-1}

\usepackage{amsfonts,amssymb,amsmath}
\usepackage[]{graphics,graphicx,epsfig}
\usepackage{amsthm}
\usepackage{color}

\def\identity{\leavevmode\hbox{\small1\kern-3.8pt\normalsize1}}

\newtheorem{propo}{Proposition}

\newcommand{\be}{\begin{eqnarray} \begin{aligned}}
\newcommand{\ee}{\end{aligned} \end{eqnarray} }
\newcommand{\bpr}{\begin{propo}}
\newcommand{\epr}{\end{propo}}
\newcommand{\bpf}{\begin{proof}}
\newcommand{\epf}{\end{proof}}
\newcommand{\ket}[1]{\left | #1 \right\rangle}
\newcommand{\bra}[1]{\left \langle #1 \right |}

\renewcommand{\t}[1]{\textrm{#1}}

\newcommand{\proj}[1]{\ket{#1}\bra{#1}}

\def\openone{\leavevmode\hbox{\small1 \normalsize \kern-.64em1}}
\renewcommand{\epsilon}{\varepsilon}

\usepackage[T1]{fontenc}

\begin{document}

%\title{Limitations on quantum computation speed-up from quantum metrological precision bounds}
\title{Quantum computation speed-up limits from quantum metrological precision bounds}

\author{Rafa\l{} Demkowicz-Dobrza\'nski}
\email{Rafal.Demkowicz-Dobrzanski@fuw.edu.pl}
\affiliation{Faculty of Physics, University of Warsaw,  ul. Pasteura 5, PL-02-093 Warszawa, Poland}
\author{Marcin Markiewicz}
\affiliation{Faculty of Physics, University of Warsaw,  ul. Pasteura 5, PL-02-093 Warszawa, Poland}
\affiliation{Institute of Theoretical Physics and Astrophysics, University of Gda\'nsk, ul. Wita Stwosza 57, 80-952 Gda\'nsk, Poland}

\begin{abstract}
We propose a  scheme for translating metrological precision bounds into lower bounds on query complexity of
quantum search algorithms. Within the scheme the link between quadratic performance enhancement in
idealized quantum metrological and quantum computing schemes becomes clear. More importantly, we utilize results from the field of quantum metrology on a generic loss of quadratic quantum precision enhancement in presence of decoherence to infer an analogous generic loss of quadratic speed-up in oracle based quantum computing. While most of our reasoning is rigorous, at one of the final steps, we need to make use of an unproven technical conjecture. We hope that we will be able to amend this deficiency in the near future, but we are convinced that even without the conjecture proven our results provide a novel and deep insight into relationship between quantum algorithms and quantum metrology protocols.
\end{abstract}

\maketitle

\section{Introduction}
Quantum metrology as well as quantum computing both aim at exploiting intrinsic quantum features such
as coherence and entanglement in order to provide enhancement over performance of corresponding classical protocols.
In a typical quantum metrological scenario \cite{Giovannetti2006, Paris2008, Banaszek2009, Giovannetti11, Demkowicz2014a, Toth2014, Kolodynski2014},
a number of probes undergo an evolution depending on an unknown parameter to be estimated. Thanks to the possibility of
preparing the probes in an entangled state or probing them sequentially in a coherent way, quantum strategies offer in principle a quadratic improvement in the scaling of precision as a function of the number of probes used. Interestingly, the quadratic improvement is also the characteristic trait of some of the oracle computational problems \cite{Bennett97, Barnum03} namely
 the Grover-type algorithms \cite{Grover96, Grover01}, in which one utilizes quantum coherence in order to reduce the number of oracle queries required to find the solution to a problem of a search through an unstructured database. Even more interestingly, studies of metrological \cite{Huelga1997, Dorner2009, Kolodynski2010, Knysh2010, Escher2011, Demkowicz12, Kolodynski2013, Knysh2014, Jarzyna2014} as well as quantum search protocols \cite{Norman99, Long00, Azuma02, Shapira03, Hsieh04, Shenvi04, Regev08, Salas08, Vrana13, Temme14} revealed that in the presence of noise the quadratic performance enhancement is
 lost in the asymptotic regime of correspondingly large number of probes or large database sizes and the quantum enhancement
 amounts to a constant factor improvement in both cases. These striking similarities call for explanation \cite{Datta12}, and making a clear connection  between the two classes of problems is the main purpose of the present paper. Additionally, since impact of decoherence on quantum protocols is much better understood in case of metrology  establishing such a connection allows one to translate general results from metrology field into quantum oracle computing field in which studies of the impact of decoherence lacked comparable generality.

In this work we show that bounds on estimation performance can be indeed directly related to the lower bounds on query complexity of quantum search algorithms by invoking limits on the speed of evolution of quantum states quantified with the help of Quantum Fisher Information (QFI). In the case of quantum parameter estimation, this fact results in the famous quantum Cram{\'e}r-Rao bound, whereas in the case of search algorithms, all known proofs of lower bounds are based on bounding the distance between states which are touched and untouched by the oracles \cite{Bennett97, Farhi98, Zalka99, Temme14}. We show, that by taking continuous-time versions of search algorithms, the proofs of their lower bounds can be expressed in the form which directly relates to QFI.  We show, that utilizing this bounds, we recover known lower bounds on noiseless quantum search, whereas in the case of noisy scenarios application of recent powerful quantum metrological methods \cite{Fujiwara2008, Escher2011, Demkowicz12} lead us to a generic conclusion that super-classical scaling of query complexity of search algorithms is asymptotically lost.

\begin{figure}
\includegraphics[width=0.9 \columnwidth]{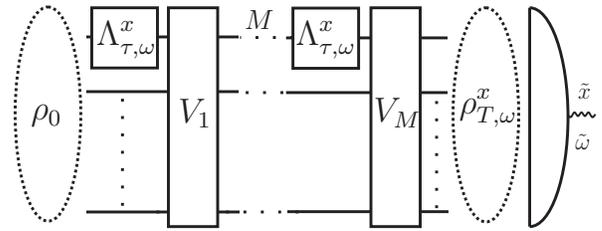}
\caption{General scheme for quantum parameter estimation and quantum search problem. Probe state $\rho_0$ is sent through a sequence of
$M$ interrogation steps $\Lambda_{\tau,\omega}^x$ each lasting time $\tau$. Finally it is measured in order to \emph{estimate} frequency parameter $\omega$ \emph{knowing} the generating Hamiltonian $x$ (metrology) or
\emph{discriminating} from an $N$ element discrete set of generating Hamiltonians $x$ \emph{knowing} the evolution frequency $\omega$ (quantum search).
Intertwining unitary operators $V_i$ guarantee full generality of the scheme covering both parallel
as well as adaptive strategies. Number of steps $M$ is arbitrary but the total interrogation time $T= M \tau$
is a fixed resource.}
\label{BlockScheme}
\end{figure}

\section{Unified scheme for parameter estimation and the quantum search problem}
Both quantum metrological as well as quantum search tasks are effectively quantum channel discrimination problems (Fig. \ref{BlockScheme}). For our purposes we will consider the quantum channel representing the interrogation step to be of the form $\Lambda_{\tau,\omega}^x(\rho) =\Lambda_{\tau}( U_{\tau,\omega}^x \rho)$,
where $U_{\tau,\omega}^x = e^{\mathrm{- i} \omega \tau H^x  }$ is the unitary sensing part of the evolution (we use a simplifying notation $U \rho  \equiv U \rho U^\dagger$), and $\Lambda_\tau$ represents undesired decoherence processes. We assume $\Lambda_\tau$ does not depend on $x$.  While $\Lambda_{\omega,\tau}^x$ acts only on one subsystem, which we will refer to as the ``sensing subsystem'', existence of ancillary particles as well as general
unitary operations $V_i$ makes this scheme completely general covering all possible sensing strategies including the adaptive ones
\cite{Giovannetti2006, Demkowicz14}. In particular when $V_i$ are chosen as operations
swapping the ``sensing subsystem'' with the $i$-th one, one arrives at a parallel sensing scenario popular in quantum metrological
literature where multiparticle input probe state $\rho_0$ is being sent through $M$ parallel channels $\Lambda_{\tau,\omega}^x$.

In the quantum metrological context, $\Lambda_{\omega,\tau}^x$
should be understood as a channel that is parameterized by an \emph{unknown} continuous frequency parameter $\omega$
that multiplies a \emph{known} Hamiltonian $H^x$ generating unitary evolution of the system for a \emph{known} time $\tau$.
The goal is to find the optimal sensing strategy that allows to infer the value of $\omega$, based on measurement results obtained from measuring the final state $\rho_{T,\omega}^x$, with minimal estimation variance under fixed total interrogation time $T=M \tau$.

In the context of quantum search algorithms, on the other hand, $\Lambda^x_{\tau,\omega}$ will represent continuous-time version of a Grover
oracle \cite{Farhi98} with an \emph{unknown} label $x$ that represents
one of the basis vectors in the $N$-dimensional Hilbert space of the ``sensing subsystem'' determining the Hamiltonian
which generates the evolution $H^x = \ket{x}\bra{x}$, where both $\omega$ and the evolution time $\tau$ are assumed to be \emph{known}.
Labels $x$ should be understood as different database entries, and for a
given database size $N$ the goal of the search problem is to determine the label $x$ under possibly minimal total interrogation time $T$.
Note an intriguing duality between the metrological and the search task. While in the first we aim at estimating the eigenvalue of the evolution operator,
in the second we aim at identifying the corresponding eigenvector. This duality makes the direct connection between the two problems less
trivial than might seem at a first sight.

For completeness let us clearly point out the relation of the continuous version of the Grover algorithm with its
original discrete version \cite{Grover96}. In the discrete version the oracle imprints a minus sign on the distinguished
eigenvector $\ket{x}$ leaving other basis vectors intact, and the goal is to determine $x$ with as few oracle calls as possible.
The original Grover algorithm requires $M \propto \sqrt{N}$ oracle calls intertwined with unitary operations
$V_i=2\proj{\psi_0}-\openone$ acting on the sensing subsystem where $\ket{\psi_0}= \frac{1}{\sqrt{N}}\sum_{x=1}^N\ket{x}$ is the initial probe state. Note that the algorithm does not utilize ancillary systems. It has been proven that up to a constant factor this is indeed
 the optimal performance even if ancillary systems where utilized under the most general strategy as depicted in Fig.~\ref{BlockScheme} \cite{Zalka99}. On the other hand, it can be easily proven that any deterministic classical strategy demands $M \propto N$ oracle calls and hence
 the quanutm speed-up is of the order of $\sqrt{N}$ in the size of the problem. In our framework we recover the original discrete scheme by setting oracle interrogation times $\tau=\pi / \omega$, which results in the total interrogation time to scale as $T \propto \sqrt{N} \pi/\omega$. In what follows, whenever one wants to relate our formulas with the discrete Grover case one should formally set $\omega=\pi$ in which case the total interrogation time $T$ will correspond to number of discrete Grover queries.

Although the performance of quantum search algorithms in the presence of noise is a subject of intensive study \cite{Norman99, Long00, Azuma02, Shapira03, Hsieh04, Shenvi04, Regev08, Salas08, Vrana13, Temme14}, it is known only for very few models of noise. Each case studied yet suggests, that arbitrarily small admixture of noise, which does not decrease with $N$, destroys the quantum speed-up, and enforces the classical asymptotic scaling of query complexity $T \propto N$. However, there exist no general characterization of the entire class of noise scenarios, which damage the super-classical scaling of query complexity of quantum search. In this work we pose a general conjecture, that this class of noise is equal to the class of noise, which  destroys the super-classical scaling of precision in quantum metrology.

\section{Quantum precision bounds for frequency estimation}
As the scheme we have introduced above involves the problem of frequency estimation let us recall known
results on the fundamental quantum precision bounds on frequency estimation in two-level systems.
Let us consider $M$ two-level atoms with transition frequency between the two levels $\ket{0}$, $\ket{1}$ equal $\omega$---a parameter to be estimated. Let the interrogation time be $\tau$ and hence in absence of decoherence and interactions each
atom evolves according to the unitary $e^{-\mathrm{i} \omega \tau \sigma_z/2}$ or
equivalently $U_{\tau,\omega}=e^{-\mathrm{i} \omega \tau \ket{1}\bra{1}}$, where $\ket{1}$ denotes the excited state.
Let $\rho_0$ be the initial state of the $M$ atoms while $\rho_{\tau,\omega} = U_{\tau,\omega}^{\otimes M} \rho_0 U_{\tau,\omega}^{\dagger \otimes M}$
be the final one. One of the main theoretical tools
to asses the limits on frequency estimation precision is the quantum Cramer-Rao bound:
\begin{equation}
\label{crb}
\delta \omega \geq \frac{1}{\sqrt{F_\omega(\rho_{\tau,\omega})}},
\end{equation}
where $F_\omega$ is the quantum Fisher information (QFI), which provided $\rho_0  =\ket{\psi_{0}}\bra{\psi_0}$
is pure, takes a simple form $F_\omega = 4 \tau^2 \Delta^2 \bar{n}$ where $\hat{n}$ is
the total excitation number operator which plays here the role of the  generating Hamiltonian, $U_{\tau,\omega}^{\otimes M}= e^{-\mathrm{i} \omega \tau \hat{n}}$. The optimal uncorrelated probe state $\ket{+}^{\otimes M}$, where $\ket{+}=(\ket{0}+\ket{1})/\sqrt{2}$,  yields  $F_\omega=\tau^2 M$, whereas using the GHZ state  $(\ket{0}^{\otimes M} + \ket{1}^{\otimes M})/\sqrt{2}$ one can boost the QFI to $F_\omega=\tau^2 M^2$ showing the quadratic precision enhancement potential.

Assessing the impact of decoherence on metrological protocols is non-trivial as the QFI needs to be calculated for mixed states
which in general requires  diagonalization of $\rho_{\tau,\omega}$ and brute force optimization of input probe states becomes
infeasible in the large $M$ limit. Fortunately, a number of theoretical tools have been developed in recent years
that allow for an efficient derivation of useful bounds \cite{Escher2011, Demkowicz12} or asymptotic formulas for QFI \cite{Knysh2014}.
These studies predict linear asymptotic scaling of the QFI with $M$ in presence of decoherence limiting the quantum correlation benefits in quantum metrology to a constant factor improvement.

To be concrete, let us consider uncorrelated Markovian model of atomic dephasing on which exposition of our approach will be based \cite{Huelga1997}.
Dephasing noise acts independently on each of the atoms, causing off-diagonal terms of atomic desity matrix,
written in $\ket{0}, \ket{1}$ basis, to be damped:
\begin{eqnarray}
\label{deph}
\Lambda_\tau&&\left( \begin{array}{cc}
\rho_{00} & \rho_{01}  \\
\rho_{10} & \rho_{11}  \end{array} \right)=
\left( \begin{array}{cc}
\rho_{00} & \rho_{01} e^{-\gamma \tau}  \\
\rho_{10} e^{-\gamma \tau} & \rho_{11}  \end{array} \right),
\end{eqnarray}
where $\gamma$ is a dephasing strength parameter. Note that this model of noise does not favour any direction in the Hilbert space. For any $\gamma > 0$ the quadratic QFI scaling is lost \cite{Huelga1997, Escher2011, Demkowicz12},
and the QFI is bounded by
\begin{equation}
\label{DephUB}
F_{\omega} \leq \tau^2 \frac{e^{-2\gamma \tau}}{1-e^{-2\gamma \tau}}M.
\end{equation}
Moreover, the bound is asymptotically saturable using e.g. one-axis spin squeezed states \cite{Orgikh2001} and therefore
represents a true limit to metrological performance in presence of dephasing. For our purposes it is important to stress that while
this bound was derived with a parallel estimation scheme in mind, it has been proven recently \cite{Demkowicz14} that it is also valid for general adaptive strategies as depicted in Fig.~\eqref{BlockScheme}.
As in typical frequency estimation approach, we treat interrogation time $\tau$
as a tunable parameter, and regard the total channel probing time
$T= \tau M$ as the resource. % and will allow us to relate this in a meaningful way with the query complexity in the search problem.
Fixing $T$ we get
\begin{equation}
\label{eq:fundamentalbound}
F_\omega\leq \tau^2 \frac{e^{-2\gamma \tau}}{1-e^{-2\gamma \tau}} \left(\frac{T}{\tau}\right) \overset{\tau \rightarrow 0}{\leq} \frac{T}{2\gamma},
\end{equation}
and the bound is maximized when we take the limit $\tau \rightarrow 0$. $F_\omega = T/2\gamma$ is therefore the ultimate saturable limit for frequency estimation in presence of dephasing, which when contrasted with the decoherence-free case  $F_\omega = \tau^2 M^2  = T^2$ reveals the quadratic enhancement loss.

\section{Limits on performance of quantum search algorithms}
In order to make use of QFI based metrological precision bounds in deriving speed-up limits of a continuous quantum search in the presence of dephasing consider as in \cite{Bennett97, Farhi98, Zalka99, Temme14} the following average probe distance quantity:
\begin{equation}
\label{dsum}
\overline{D}_{T}=\sum_x D(\rho_{T,\omega}^x,\rho_{T}),
\end{equation}
where $D(\rho_1,\rho_2)$ is a distance measure to be specified below, $\rho_{T,\omega}^x$ is the final state of the algorithm, whereas $\rho_{T}$ is the state of the same algorithm (the same set of unitaries $V_i$), but with the unitary sensing part removed from the oracle queries i.e. $\Lambda_{\tau,\omega}^x$ is replaced by the decoherence map $\Lambda_\tau$, or equivalently $\omega=0$ is set. In order to make the connection to metrological bounds, we choose as the distance measure the angular Bures distance \cite{Bures69, Helstrom67,Bengtsson2006}:
\begin{equation}
\label{Bures}
D(\rho_1,\rho_2)=\arccos\left(\t{Tr}\sqrt{\rho_1^{1/2}\rho_2 \rho_1^{1/2}}\right),
\end{equation}
which for infinitesimally close states lying along a trajectory parameterized by a continuous parameter is expressed in terms of the QFI:
$D(\rho_\omega, \rho_{\omega+\t{d}\omega}) =\frac{1}{2}\sqrt{F_\omega(\rho_\omega)}\, \t{d}\omega $.
From a geometrical point of view, QFI introduces a natural metric in the space of density matrices \cite{Braunstein94}---the Bures metric---and
as a consequence a measure of speed of evolution of quantum states \cite{Giovannetti03, Taddei13}.
This geometric context is the core of the proposed correspondence between performance of parameter estimation and query complexity of search problems.

We aim at deriving a lower bound on the total interrogation time $T$ necessary for the search problem to succeed, in presence of dephasing noise $\Lambda_\tau$ which is taken as a natural generalization of the two-level dephasing map to $N$ level systems on which Grover query acts, namely a map which does not distinguish any of the basis states $\ket{1},\dots,\ket{N}$ and simply damps all of the off-diagonal terms with the same damping coefficient $\exp(-\gamma \tau)$. In order for the search algorithm to succeed the final states
$\rho^{x}_{T,\omega}$ should be perfectly distinguishable, i.e. occupy orthogonal subspaces for different labels $x$.
%(for simplicity we assume success probability $1$, though finite success probability might be incorporated in the exposition without qualitative %changes).
This implies
\begin{equation}
\label{dub}
\forall_{x\neq x'}D(\rho_{T,\omega}^x,\rho_{T,\omega}^{x'})=\pi/2.
\end{equation}
Using the triangle inequality we have $D(\rho_{T,\omega}^x,\rho_T)+D(\rho_T,\rho_{T,\omega}^{x'})\geq \pi/2$, hence we obtain the lower bound:
\begin{equation}
\label{dtLB}
\overline{D}_T\geq N\frac{\pi}{4}.
\end{equation}

\begin{figure}
\includegraphics[width=0.9 \columnwidth]{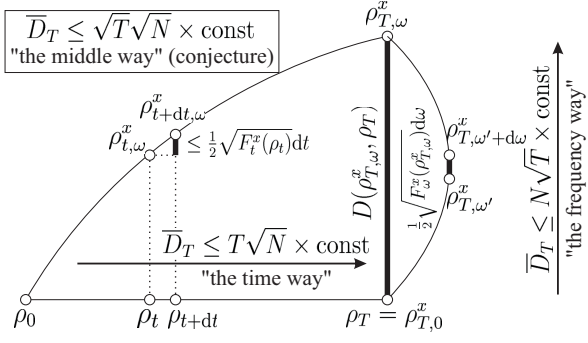}
\caption{Schematic representation of proofs of upper bounds on the probe states
average distance $\overline{D}$ obtained by either summing distance increments over time or frequency.}
\label{fig:paths}
\end{figure}
We will now present two ways leading to upper bounds on $\overline{D}_T$ which will eventually allow us to draw conclusions on minimal interrogation time $T$ required in the search problem. First, ``the time way'' is indicated in Fig.~\ref{fig:paths} by a horizontal arrow.
Let $\rho_{t,\omega}^x$, $\rho_t$ denote the probe state at some intermediate time $t$ (that is after total number of $t/\tau$ queries)
evolving respectively in presence or in the absence of the unitary part of the query. Consider the following chain of (in)equalities:
\begin{eqnarray}
\label{MainIneq}
&&D(\rho_{t+\textrm{d}t,\omega}^x,\rho_{t+\textrm{d}t})=D[V \Lambda_{\t{d} t}( U^x_{\t{d} t,\omega}\rho_{t,\omega}^x) ,
V \Lambda_{\t{d} t}( \rho_{t})]=\nonumber\\
&&=D[\Lambda_{\t{d} t}( U^x_{\t{d} t,\omega}\rho_{t,\omega}^x) ,
\Lambda_{\t{d} t}( \rho_{t})]\leq  D(U^x_{\t{d} t,\omega}\rho_{t,\omega}^x ,
\rho_{t}) = \nonumber\\
&&=  D(\rho_{t,\omega}^x,U^{x \dagger}_{\t{d} t,\omega} \rho_t) \leq  D(\rho_{t,\omega}^x,\rho_{t})+D(\rho_t,U^{x \dagger}_{\t{d}t,\omega}\rho_{t}),
\end{eqnarray}
where we have made use of invariance of Bures distance under unitary transformations, nonincreasing of the Bures distance
under general quantum maps and finally the triangle inequality ($V$ represents intertwining unitary operation at time $t$).
We can therefore bound the increase of distance between the states as:
\begin{equation}
D(\rho_{t+\t{d}t,\omega}^x,\rho_{t + \t{d}t})-D(\rho_{t,\omega}^x,\rho_{t})\leq D(\rho_t,U^{x \dagger}_{\t{d}t,\omega}\rho_{t}).
\end{equation}
Invoking the relation between infinitesimal Bures distances and the QFI we get:
\begin{equation}
\label{dtauD}
%\frac{\operatorname d}{\operatorname dt}D(\rho_{t,\omega}^x,\rho_{t})\leq \frac{1}{2}\sqrt{F_t^x(\rho_t)},
\frac{\operatorname d \overline{D}_t}{\operatorname dt}\leq \frac{1}{2} \sum_x \sqrt{F_t^x(\rho_t)},
\end{equation}
where $F_t^x$ represents QFI with respect to a unitary time evolution parameter with generating Hamiltonian $\omega \ket{x}\bra{x}$.
Since in general $\rho_t$ is mixed, QFI is not proportional to the variance of the generator as in the pure state case.
Still, when considering all convex decompositions $\rho_t = \sum_i p_i \ket{\psi_i}\bra{\psi_i}$, $p_i\geq 0$,
it is known that the QFI of $\rho_t$ is equal to the minimum
of weighted sum of QFIs over all decompositions \cite{Toth2013, Yu2013}:
\begin{eqnarray}
F_t^x(\rho_{t})&&=\min_{\{p_i,\proj{\psi_i}\}}\sum_i p_i 4 \Delta^2( \omega \proj{x})_{\ket{\psi_i}} \nonumber \\
&&\leq \sum_i p_i \bra{\psi_i} (\omega \proj{x})^2 \ket{\psi_i},
\end{eqnarray}
where the upper bound follows from replacing the variance by expectation of the squared generator and a particular decomposition $\ket{\psi_i}$ (e.g. eigendecomposition) is chosen.
Now it follows that:
\begin{equation}
\sum_x F^x_t(\rho_{\tau})\leq 4 \omega^2\sum_i p_i\bra{\psi_i}\left(\sum_x \proj{x}\right)\ket{\psi_i}=4 \omega^2.
\end{equation}
As in \cite{Farhi98} we use the property: $\sum_{i=1}^N |a_i|^2=1\Longrightarrow \sum_{i=1}^N |a_i|\leq\sqrt{N}$ to get:
\begin{equation}
\label{sqrtF}
\sum_x \sqrt{F^x_t(\rho_t)}\leq 2\omega \sqrt{N}.
\end{equation}
Combining \eqref{dtauD} and \eqref{sqrtF} we get
\be
\label{noiselessUB}
\overline{D}_{T}\leq T \sqrt{N}\omega,
\ee
which together with the lower bound \eqref{dtLB} yields
\be
\label{noiselessQC}
T\geq \frac{\pi}{4\omega}\sqrt{N}.
\ee
This is a known bound on the performance of ideal, decoherence-free Grover algorithm \cite{Farhi98}. It is therefore clear that
by studying the Bures distance growth over time alone we are not able to capture the essence of the impact of decoherence
on the algorithm performance. The reason for this is that integration over time leading to  \eqref{noiselessUB} yielded the bound on $\overline{D}_T$ which scales linearly with $T$ which is not the true scaling of $\overline{D}_T$ as will be demonstrated below.

We now take ``the frequency way'' approach which is depicted by a vertical arrow in Fig.~\ref{fig:paths}, and study the growth of
the distance $D(\rho^x_{T,\omega^\prime},\rho_T)$  while increasing frequency parameter $\omega^\prime$ from  $0$ to $\omega$.
Note that $\rho^x_{T,0}=\rho_T$ so the initial distance is $0$.  Using the triangle inequality as well as the fundamental bound
on QFI for frequency estimation in presence of dephasing \eqref{eq:fundamentalbound} we get:
\begin{multline}
\label{mainInt}
D(\rho_{T,\omega}^{x},\rho_T)\leq \int_0^{\omega}D(\rho_{T,\omega^\prime+\t{d}\omega}^x,\rho_{T,\omega^\prime}^{x})= \nonumber\\
 = \frac{1}{2} \int_0^{\omega} \sqrt{F_{\omega^\prime}^x(\rho_{T,\omega^\prime}^x)} \t{d}\omega^\prime \leq \frac{\omega}{2\sqrt{2 \gamma}}\sqrt{T},
\end{multline}
which clearly demonstrates that the time scaling of $\overline{D}_T$ is bounded by $\sqrt{T}$. The idea of integrating over frequency rather than time is  the crux of the paper as it allows to link metrological bounds with a quantity useful to asses the performance of the search algorithm. We should stress that analogous bounds could be derived for an arbitrary decoherence model in which QFI scaling is linear in $T$, which is a generic case when uncorrelated noise is present in the system \cite{Demkowicz12, Kolodynski2013, Knysh2014}. Let us also remind that in absence of decoherence
QFI scales as $T^2$ allowing for a linear time increase of $\overline{D}_T$ and as a result the quadratic speed-up in noiseless Grover algorithm.
In order to calculate $\overline{D}_T$ we now need to
find the bound on the $\sum_x D(\rho_{T,\omega}^{x},\rho_T)$. For this we need to bound
$\sum_x \sqrt{F_\omega^x(\rho_{T,\omega}^x)}$. In the case of ``the time way'' derivation we were able to show that this sum scales as $\sqrt{N}$.
Intuitively, this is due to the fact that there are no probe states that would be optimally sensitive to all unitary evolutions with
different generators $\proj{x}$ that form a basis in the sensing Hilbert space. As the sensing evolution $x$ imprints the phase only
on one basis vector $\ket{x}$, a state which aims at optimizing the sum of QFIs for all $x$ need to have non-zero overlap with all the basis states. This makes an individual QFI for a particular $x$ smaller than it in principle could be if we designed the optimal probing state
for frequency estimation \emph{knowing} $x$. We expect that this will again cause $\sum_x \sqrt{F_\omega^x(\rho_{T,\omega}^x)}$ to scale like
$\sqrt{N}$. Still, due to the very general structure of the scheme we consider, which in principle involves adaptive strategies, we were not able to rigorously prove this fact. Therefore we perform a trivial sum over $x$ and arrive at:
\begin{equation}
\label{eq:boundfreq}
\overline{D}_T \leq \frac{\omega}{2\sqrt{2 \gamma}} N \sqrt{T}.
\end{equation}
Inspecting \eqref{noiselessUB} and \eqref{eq:boundfreq} we conclude that
for fixed $T$ the quantity $\overline{D}_T$ cannot grow faster that $\sqrt{N}$ and on the other hand for a fixed $N$
it cannot grow faster than $\sqrt{T}$. Unfortunately, without additional technical assumptions
on the properties of $\overline{D}_T$, e.g. that asymptotically $\overline{D}_T \propto N^\alpha T^\beta$,
we cannot rigorously conclude that
\begin{equation}
\label{eq:conjecture}
\overline{D}_T \leq \sqrt{T} \sqrt{N} \times \t{const}.
\end{equation}
We therefore need to leave the above formula as a natural conjecture that arises from our reasonings, and hope that
an approach that would exploit the advantages of both the time and the frequency approaches will be capable of proving the above conjecture rigorously.
Combining \eqref{eq:conjecture} with \eqref{dtLB} we arrive at the desired result:
\begin{equation}
T\geq \t{const} \times N,
\end{equation}
that the quadratic quantum enhancement in the search algorithm is lost.

In order to confirm our conjecture at least numerically, we have performed a numerical optimization
of $\sum_x \sqrt{F_\omega^x(\rho_{T,\omega}^x)}$ in a restricted basic parallel estimation scheme that does not
allow for adaptive probe state transformations. We were able to numerically confirm that
 $ \max_{\rho_0} \sum_x \sqrt{F_\omega^x(\rho_{t,\omega}^x)} \leq 2\sqrt{N} \max_{\rho_0} \sqrt{F_\omega^{x_0}(\rho_{t,\omega}^{x_0})}$, where on the right hand side of the inequality we assume that we know the generating Hamiltonian $x=x_0$. Keeping track of the constants we can derive a quantitative bound $T\geq N\left(\frac{\pi^2}{8}\right)\frac{\gamma}{\omega^2}$. This bound has a similar structure to the lower bound on the complexity of a continuous quantum search in presence of dephasing in a sequential scenario without ancillas, proposed in \cite{Temme14}: $T\geq N\left(\frac{2\gamma}{\gamma^2+4\omega^2}\right)$. Note that for fixed $N$ our bound is larger (and therefore more restrictive) for any set of parameters $\omega$ and $\gamma$.

\section{Conclusions}
We proposed a general scheme for translating metrological precision bounds in terms of QFI into lower bounds on query complexity of quantum search problems. This allowed us to derive the lower bound on query complexity of quantum search directly from the properties of QFI. We posed a conjecture, which assumes a non-trivial upper bound on the sum of square-roots of the QFIs for frequency estimation around different orthogonal axes. The validity of the conjecture implies, that any model of noise, which suppresses the quantum gain in estimation precision in the most general estimation scenario, also destroys the quantum speed-up in quantum search algorithms. 
%To the best of our knowledge our framework provides the first direct link between metrological precision bounds and computational lower bounds.

A natural continuation of presented work would be finding the connection between nonlinear versions of search algorithms \cite{Meyer14} and nonlinear metrological scenarios \cite{Datta12}. On the one hand such research can shed a new light on the problem of computational power of nonlinear quantum evolution, on the other it can stimulate the development of noisy nonlinear metrological schemes in the most general scenarios.
%If our conjecture is valid, the problem of characterizing models of noise which destroy quantum speed-up in query complexity is equivalent to %characterizing models of noise, which destroy super-classical scaling of estimation precision in the general estimation scenario.

\section*{Acknowledgements} We would like to thank Janek Ko{\l}odynski, Christian Gogolin, Marcin Jarzyna and Marcin Zwierz  for a number of
inspiring discussions and attempts to prove the conjecture. This research was supported by the EC under the FP7 IP project SIQS co-financed by the Polish Ministry of Science and Higher Education and by the Polish Ministry of Science and Higher Education ``Iuventus'' Plus program for years 2015-2017 No. 0088/IP3/2015/73.

\bibliographystyle{apsrev4-1}
\bibliography{Grover}

\begin{thebibliography}{45}%
\makeatletter
\providecommand \@ifxundefined [1]{%
 \@ifx{#1\undefined}
}%
\providecommand \@ifnum [1]{%
 \ifnum #1\expandafter \@firstoftwo
 \else \expandafter \@secondoftwo
 \fi
}%
\providecommand \@ifx [1]{%
 \ifx #1\expandafter \@firstoftwo
 \else \expandafter \@secondoftwo
 \fi
}%
\providecommand \natexlab [1]{#1}%
\providecommand \enquote  [1]{``#1''}%
\providecommand \bibnamefont  [1]{#1}%
\providecommand \bibfnamefont [1]{#1}%
\providecommand \citenamefont [1]{#1}%
\providecommand \href@noop [0]{\@secondoftwo}%
\providecommand \href [0]{\begingroup \@sanitize@url \@href}%
\providecommand \@href[1]{\@@startlink{#1}\@@href}%
\providecommand \@@href[1]{\endgroup#1\@@endlink}%
\providecommand \@sanitize@url [0]{\catcode `\\12\catcode `\$12\catcode
  `\&12\catcode `\#12\catcode `\^12\catcode `\_12\catcode `\%12\relax}%
\providecommand \@@startlink[1]{}%
\providecommand \@@endlink[0]{}%
\providecommand \url  [0]{\begingroup\@sanitize@url \@url }%
\providecommand \@url [1]{\endgroup\@href {#1}{\urlprefix }}%
\providecommand \urlprefix  [0]{URL }%
\providecommand \Eprint [0]{\href }%
\providecommand \doibase [0]{http://dx.doi.org/}%
\providecommand \selectlanguage [0]{\@gobble}%
\providecommand \bibinfo  [0]{\@secondoftwo}%
\providecommand \bibfield  [0]{\@secondoftwo}%
\providecommand \translation [1]{[#1]}%
\providecommand \BibitemOpen [0]{}%
\providecommand \bibitemStop [0]{}%
\providecommand \bibitemNoStop [0]{.\EOS\space}%
\providecommand \EOS [0]{\spacefactor3000\relax}%
\providecommand \BibitemShut  [1]{\csname bibitem#1\endcsname}%
\let\auto@bib@innerbib\@empty
%</preamble>
\bibitem [{\citenamefont {Giovannetti}\ \emph {et~al.}(2006)\citenamefont
  {Giovannetti}, \citenamefont {Lloyd},\ and\ \citenamefont
  {Maccone}}]{Giovannetti2006}%
  \BibitemOpen
  \bibfield  {author} {\bibinfo {author} {\bibfnamefont {V.}~\bibnamefont
  {Giovannetti}}, \bibinfo {author} {\bibfnamefont {S.}~\bibnamefont {Lloyd}},
  \ and\ \bibinfo {author} {\bibfnamefont {L.}~\bibnamefont {Maccone}},\ }\href
  {\doibase 10.1103/PhysRevLett.96.010401} {\bibfield  {journal} {\bibinfo
  {journal} {Phys. Rev. Lett.}\ }\textbf {\bibinfo {volume} {96}},\ \bibinfo
  {pages} {010401} (\bibinfo {year} {2006})}\BibitemShut {NoStop}%
\bibitem [{\citenamefont {Paris}(2009)}]{Paris2008}%
  \BibitemOpen
  \bibfield  {author} {\bibinfo {author} {\bibfnamefont {M.~G.~A.}\
  \bibnamefont {Paris}},\ }\href {http://arxiv.org/abs/0804.2981} {\bibfield
  {journal} {\bibinfo  {journal} {Int. J. Quantum Inf.}\ }\textbf {\bibinfo
  {volume} {7}},\ \bibinfo {pages} {125} (\bibinfo {year} {2009})}\BibitemShut
  {NoStop}%
\bibitem [{\citenamefont {Banaszek}\ \emph {et~al.}(2009)\citenamefont
  {Banaszek}, \citenamefont {Demkowicz-Dobrzanski},\ and\ \citenamefont
  {Walmsley}}]{Banaszek2009}%
  \BibitemOpen
  \bibfield  {author} {\bibinfo {author} {\bibfnamefont {K.}~\bibnamefont
  {Banaszek}}, \bibinfo {author} {\bibfnamefont {R.}~\bibnamefont
  {Demkowicz-Dobrzanski}}, \ and\ \bibinfo {author} {\bibfnamefont {I.~A.}\
  \bibnamefont {Walmsley}},\ }\href@noop {} {\bibfield  {journal} {\bibinfo
  {journal} {Nature Photon.}\ }\textbf {\bibinfo {volume} {3}},\ \bibinfo
  {pages} {673} (\bibinfo {year} {2009})}\BibitemShut {NoStop}%
\bibitem [{\citenamefont {V.~Giovannetti}\ and\ \citenamefont
  {Maccone}(2011)}]{Giovannetti11}%
  \BibitemOpen
  \bibfield  {author} {\bibinfo {author} {\bibfnamefont {S.~L.}\ \bibnamefont
  {V.~Giovannetti}}\ and\ \bibinfo {author} {\bibfnamefont {L.}~\bibnamefont
  {Maccone}},\ }\href@noop {} {\bibfield  {journal} {\bibinfo  {journal}
  {Nature Phot.}\ }\textbf {\bibinfo {volume} {5}},\ \bibinfo {pages} {222}
  (\bibinfo {year} {2011})}\BibitemShut {NoStop}%
\bibitem [{\citenamefont {{Demkowicz-Dobrzanski}}\ \emph
  {et~al.}(2014)\citenamefont {{Demkowicz-Dobrzanski}}, \citenamefont
  {{Jarzyna}},\ and\ \citenamefont {{Kolodynski}}}]{Demkowicz2014a}%
  \BibitemOpen
  \bibfield  {author} {\bibinfo {author} {\bibfnamefont {R.}~\bibnamefont
  {{Demkowicz-Dobrzanski}}}, \bibinfo {author} {\bibfnamefont {M.}~\bibnamefont
  {{Jarzyna}}}, \ and\ \bibinfo {author} {\bibfnamefont {J.}~\bibnamefont
  {{Kolodynski}}},\ }\href@noop {} {\bibfield  {journal} {\bibinfo  {journal}
  {ArXiv e-prints}\ } (\bibinfo {year} {2014})},\ \Eprint
  {http://arxiv.org/abs/1405.7703} {arXiv:1405.7703 [quant-ph]} \BibitemShut
  {NoStop}%
\bibitem [{\citenamefont {Tóth}\ and\ \citenamefont
  {Apellaniz}(2014)}]{Toth2014}%
  \BibitemOpen
  \bibfield  {author} {\bibinfo {author} {\bibfnamefont {G.}~\bibnamefont
  {Tóth}}\ and\ \bibinfo {author} {\bibfnamefont {I.}~\bibnamefont
  {Apellaniz}},\ }\href {http://stacks.iop.org/1751-8121/47/i=42/a=424006}
  {\bibfield  {journal} {\bibinfo  {journal} {Journal of Physics A:
  Mathematical and Theoretical}\ }\textbf {\bibinfo {volume} {47}},\ \bibinfo
  {pages} {424006} (\bibinfo {year} {2014})}\BibitemShut {NoStop}%
\bibitem [{\citenamefont {{Kolodynski}}(2014)}]{Kolodynski2014}%
  \BibitemOpen
  \bibfield  {author} {\bibinfo {author} {\bibfnamefont {J.}~\bibnamefont
  {{Kolodynski}}},\ }\emph {\bibinfo {title} {{Precision bounds in noisy
  quantum metrology}}},\ \href@noop {} {Ph.D. thesis} (\bibinfo {year}
  {2014}),\ \Eprint {http://arxiv.org/abs/1409.0535} {arXiv:1409.0535
  [quant-ph]} \BibitemShut {NoStop}%
\bibitem [{\citenamefont {Bennett}\ \emph {et~al.}(1997)\citenamefont
  {Bennett}, \citenamefont {Bernstein}, \citenamefont {Brassard},\ and\
  \citenamefont {Vazirani}}]{Bennett97}%
  \BibitemOpen
  \bibfield  {author} {\bibinfo {author} {\bibfnamefont {C.~H.}\ \bibnamefont
  {Bennett}}, \bibinfo {author} {\bibfnamefont {E.}~\bibnamefont {Bernstein}},
  \bibinfo {author} {\bibfnamefont {G.}~\bibnamefont {Brassard}}, \ and\
  \bibinfo {author} {\bibfnamefont {U.}~\bibnamefont {Vazirani}},\ }\href@noop
  {} {\bibfield  {journal} {\bibinfo  {journal} {SIAM J. Comput.}\ }\textbf
  {\bibinfo {volume} {26}},\ \bibinfo {pages} {1510} (\bibinfo {year}
  {1997})}\BibitemShut {NoStop}%
\bibitem [{\citenamefont {Barnum}\ \emph {et~al.}(2003)\citenamefont {Barnum},
  \citenamefont {Saks},\ and\ \citenamefont {Szegedy}}]{Barnum03}%
  \BibitemOpen
  \bibfield  {author} {\bibinfo {author} {\bibfnamefont {H.}~\bibnamefont
  {Barnum}}, \bibinfo {author} {\bibfnamefont {H.}~\bibnamefont {Saks}}, \ and\
  \bibinfo {author} {\bibfnamefont {M.}~\bibnamefont {Szegedy}},\ }\href@noop
  {} {\bibfield  {journal} {\bibinfo  {journal} {Proceedings of 18th IEEE
  Annual Conference on Computational Complexity}\ ,\ \bibinfo {pages} {179}}
  (\bibinfo {year} {2003})}\BibitemShut {NoStop}%
\bibitem [{\citenamefont {Grover}(1996)}]{Grover96}%
  \BibitemOpen
  \bibfield  {author} {\bibinfo {author} {\bibfnamefont {L.~K.}\ \bibnamefont
  {Grover}},\ }\href@noop {} {\bibfield  {journal} {\bibinfo  {journal} {Proc.
  28th Annual ACM STOC}\ ,\ \bibinfo {pages} {212}} (\bibinfo {year}
  {1996})}\BibitemShut {NoStop}%
\bibitem [{\citenamefont {Grover}(2001)}]{Grover01}%
  \BibitemOpen
  \bibfield  {author} {\bibinfo {author} {\bibfnamefont {L.~K.}\ \bibnamefont
  {Grover}},\ }\href@noop {} {\bibfield  {journal} {\bibinfo  {journal} {Am. J.
  Phys.}\ }\textbf {\bibinfo {volume} {69}},\ \bibinfo {pages} {769} (\bibinfo
  {year} {2001})}\BibitemShut {NoStop}%
\bibitem [{\citenamefont {Huelga}\ \emph {et~al.}(1997)\citenamefont {Huelga},
  \citenamefont {Macchiavello}, \citenamefont {Pellizzari}, \citenamefont
  {Ekert}, \citenamefont {Plenio},\ and\ \citenamefont {Cirac}}]{Huelga1997}%
  \BibitemOpen
  \bibfield  {author} {\bibinfo {author} {\bibfnamefont {S.~F.}\ \bibnamefont
  {Huelga}}, \bibinfo {author} {\bibfnamefont {C.}~\bibnamefont
  {Macchiavello}}, \bibinfo {author} {\bibfnamefont {T.}~\bibnamefont
  {Pellizzari}}, \bibinfo {author} {\bibfnamefont {A.~K.}\ \bibnamefont
  {Ekert}}, \bibinfo {author} {\bibfnamefont {M.~B.}\ \bibnamefont {Plenio}}, \
  and\ \bibinfo {author} {\bibfnamefont {J.~I.}\ \bibnamefont {Cirac}},\ }\href
  {\doibase 10.1103/PhysRevLett.79.3865} {\bibfield  {journal} {\bibinfo
  {journal} {Phys. Rev. Lett.}\ }\textbf {\bibinfo {volume} {79}},\ \bibinfo
  {pages} {3865} (\bibinfo {year} {1997})}\BibitemShut {NoStop}%
\bibitem [{\citenamefont {Dorner}\ \emph {et~al.}(2009)\citenamefont {Dorner},
  \citenamefont {Demkowicz-Dobrza{\'n}ski}, \citenamefont {Smith},
  \citenamefont {Lundeen}, \citenamefont {Wasilewski}, \citenamefont
  {Banaszek},\ and\ \citenamefont {Walmsley}}]{Dorner2009}%
  \BibitemOpen
  \bibfield  {author} {\bibinfo {author} {\bibfnamefont {U.}~\bibnamefont
  {Dorner}}, \bibinfo {author} {\bibfnamefont {R.}~\bibnamefont
  {Demkowicz-Dobrza{\'n}ski}}, \bibinfo {author} {\bibfnamefont {B.~J.}\
  \bibnamefont {Smith}}, \bibinfo {author} {\bibfnamefont {J.~S.}\ \bibnamefont
  {Lundeen}}, \bibinfo {author} {\bibfnamefont {W.}~\bibnamefont {Wasilewski}},
  \bibinfo {author} {\bibfnamefont {K.}~\bibnamefont {Banaszek}}, \ and\
  \bibinfo {author} {\bibfnamefont {I.~A.}\ \bibnamefont {Walmsley}},\ }\href
  {\doibase 10.1103/PhysRevLett.102.040403} {\bibfield  {journal} {\bibinfo
  {journal} {Phys. Rev. Lett.}\ }\textbf {\bibinfo {volume} {102}},\ \bibinfo
  {pages} {040403} (\bibinfo {year} {2009})}\BibitemShut {NoStop}%
\bibitem [{\citenamefont {Ko\l{}ody\'{n}ski}\ and\ \citenamefont
  {Demkowicz-Dobrza\'{n}ski}(2010)}]{Kolodynski2010}%
  \BibitemOpen
  \bibfield  {author} {\bibinfo {author} {\bibfnamefont {J.}~\bibnamefont
  {Ko\l{}ody\'{n}ski}}\ and\ \bibinfo {author} {\bibfnamefont {R.}~\bibnamefont
  {Demkowicz-Dobrza\'{n}ski}},\ }\href {\doibase 10.1103/PhysRevA.82.053804}
  {\bibfield  {journal} {\bibinfo  {journal} {Phys. Rev. A}\ }\textbf {\bibinfo
  {volume} {82}},\ \bibinfo {pages} {053804} (\bibinfo {year}
  {2010})}\BibitemShut {NoStop}%
\bibitem [{\citenamefont {Knysh}\ \emph {et~al.}(2011)\citenamefont {Knysh},
  \citenamefont {Smelyanskiy},\ and\ \citenamefont {Durkin}}]{Knysh2010}%
  \BibitemOpen
  \bibfield  {author} {\bibinfo {author} {\bibfnamefont {S.}~\bibnamefont
  {Knysh}}, \bibinfo {author} {\bibfnamefont {V.~N.}\ \bibnamefont
  {Smelyanskiy}}, \ and\ \bibinfo {author} {\bibfnamefont {G.~A.}\ \bibnamefont
  {Durkin}},\ }\href {\doibase 10.1103/PhysRevA.83.021804} {\bibfield
  {journal} {\bibinfo  {journal} {Phys. Rev. A}\ }\textbf {\bibinfo {volume}
  {83}},\ \bibinfo {pages} {021804} (\bibinfo {year} {2011})}\BibitemShut
  {NoStop}%
\bibitem [{\citenamefont {Escher}\ \emph {et~al.}(2011)\citenamefont {Escher},
  \citenamefont {de~Matos~Filho},\ and\ \citenamefont
  {Davidovich}}]{Escher2011}%
  \BibitemOpen
  \bibfield  {author} {\bibinfo {author} {\bibfnamefont {B.~M.}\ \bibnamefont
  {Escher}}, \bibinfo {author} {\bibfnamefont {R.~L.}\ \bibnamefont
  {de~Matos~Filho}}, \ and\ \bibinfo {author} {\bibfnamefont {L.}~\bibnamefont
  {Davidovich}},\ }\href {\doibase 10.1038/nphys1958} {\bibfield  {journal}
  {\bibinfo  {journal} {Nature Phys.}\ }\textbf {\bibinfo {volume} {7}},\
  \bibinfo {pages} {406} (\bibinfo {year} {2011})}\BibitemShut {NoStop}%
\bibitem [{\citenamefont {R.~Demkowicz-Dobrzanski}\ and\ \citenamefont
  {Guta}(2012)}]{Demkowicz12}%
  \BibitemOpen
  \bibfield  {author} {\bibinfo {author} {\bibfnamefont {J.~K.}\ \bibnamefont
  {R.~Demkowicz-Dobrzanski}}\ and\ \bibinfo {author} {\bibfnamefont
  {M.}~\bibnamefont {Guta}},\ }\href@noop {} {\bibfield  {journal} {\bibinfo
  {journal} {Nature Communications}\ }\textbf {\bibinfo {volume} {3}},\
  \bibinfo {pages} {1063} (\bibinfo {year} {2012})}\BibitemShut {NoStop}%
\bibitem [{\citenamefont {Ko{\l}ody{\'n}ski}\ and\ \citenamefont
  {Demkowicz-Dobrza{\'n}ski}(2013)}]{Kolodynski2013}%
  \BibitemOpen
  \bibfield  {author} {\bibinfo {author} {\bibfnamefont {J.}~\bibnamefont
  {Ko{\l}ody{\'n}ski}}\ and\ \bibinfo {author} {\bibfnamefont {R.}~\bibnamefont
  {Demkowicz-Dobrza{\'n}ski}},\ }\href {\doibase 10.1088/1367-2630/15/7/073043}
  {\bibfield  {journal} {\bibinfo  {journal} {New J. Phys.}\ }\textbf {\bibinfo
  {volume} {15}},\ \bibinfo {pages} {073043} (\bibinfo {year}
  {2013})}\BibitemShut {NoStop}%
\bibitem [{\citenamefont {Knysh}\ \emph {et~al.}(2014)\citenamefont {Knysh},
  \citenamefont {Chen},\ and\ \citenamefont {Durkin}}]{Knysh2014}%
  \BibitemOpen
  \bibfield  {author} {\bibinfo {author} {\bibfnamefont {S.~I.}\ \bibnamefont
  {Knysh}}, \bibinfo {author} {\bibfnamefont {E.~H.}\ \bibnamefont {Chen}}, \
  and\ \bibinfo {author} {\bibfnamefont {G.~A.}\ \bibnamefont {Durkin}},\
  }\href@noop {} {\bibfield  {journal} {\bibinfo  {journal} {ArXiv e-prints}\ }
  (\bibinfo {year} {2014})},\ \Eprint {http://arxiv.org/abs/1402.0495}
  {arXiv:1402.0495 [quant-ph]} \BibitemShut {NoStop}%
\bibitem [{\citenamefont {{Jarzyna}}\ and\ \citenamefont
  {{Demkowicz-Dobrzanski}}(2014)}]{Jarzyna2014}%
  \BibitemOpen
  \bibfield  {author} {\bibinfo {author} {\bibfnamefont {M.}~\bibnamefont
  {{Jarzyna}}}\ and\ \bibinfo {author} {\bibfnamefont {R.}~\bibnamefont
  {{Demkowicz-Dobrzanski}}},\ }\href@noop {} {\bibfield  {journal} {\bibinfo
  {journal} {ArXiv e-prints}\ } (\bibinfo {year} {2014})},\ \Eprint
  {http://arxiv.org/abs/1407.4805} {arXiv:1407.4805 [quant-ph]} \BibitemShut
  {NoStop}%
\bibitem [{\citenamefont {Pablo-Norman}\ and\ \citenamefont
  {Ruiz-Altaba}(1999)}]{Norman99}%
  \BibitemOpen
  \bibfield  {author} {\bibinfo {author} {\bibfnamefont {B.}~\bibnamefont
  {Pablo-Norman}}\ and\ \bibinfo {author} {\bibfnamefont {M.}~\bibnamefont
  {Ruiz-Altaba}},\ }\href@noop {} {\bibfield  {journal} {\bibinfo  {journal}
  {Phys. Rev. A}\ }\textbf {\bibinfo {volume} {61}},\ \bibinfo {pages} {012301}
  (\bibinfo {year} {1999})}\BibitemShut {NoStop}%
\bibitem [{\citenamefont {Long}\ \emph {et~al.}(2000)\citenamefont {Long},
  \citenamefont {Li}, \citenamefont {Zhang},\ and\ \citenamefont
  {Tu}}]{Long00}%
  \BibitemOpen
  \bibfield  {author} {\bibinfo {author} {\bibfnamefont {G.}~\bibnamefont
  {Long}}, \bibinfo {author} {\bibfnamefont {Y.}~\bibnamefont {Li}}, \bibinfo
  {author} {\bibfnamefont {W.}~\bibnamefont {Zhang}}, \ and\ \bibinfo {author}
  {\bibfnamefont {C.}~\bibnamefont {Tu}},\ }\href@noop {} {\bibfield  {journal}
  {\bibinfo  {journal} {Phys. Rev. A}\ }\textbf {\bibinfo {volume} {61}},\
  \bibinfo {pages} {042305} (\bibinfo {year} {2000})}\BibitemShut {NoStop}%
\bibitem [{\citenamefont {Azuma}(2002)}]{Azuma02}%
  \BibitemOpen
  \bibfield  {author} {\bibinfo {author} {\bibfnamefont {H.}~\bibnamefont
  {Azuma}},\ }\href@noop {} {\bibfield  {journal} {\bibinfo  {journal} {Phys.
  Rev. A}\ }\textbf {\bibinfo {volume} {65}},\ \bibinfo {pages} {042311}
  (\bibinfo {year} {2002})}\BibitemShut {NoStop}%
\bibitem [{\citenamefont {Shapira}\ \emph {et~al.}(2003)\citenamefont
  {Shapira}, \citenamefont {Mozes},\ and\ \citenamefont {Biham}}]{Shapira03}%
  \BibitemOpen
  \bibfield  {author} {\bibinfo {author} {\bibfnamefont {D.}~\bibnamefont
  {Shapira}}, \bibinfo {author} {\bibfnamefont {S.}~\bibnamefont {Mozes}}, \
  and\ \bibinfo {author} {\bibfnamefont {O.}~\bibnamefont {Biham}},\
  }\href@noop {} {\bibfield  {journal} {\bibinfo  {journal} {Phys. Rev. A}\
  }\textbf {\bibinfo {volume} {67}},\ \bibinfo {pages} {042301} (\bibinfo
  {year} {2003})}\BibitemShut {NoStop}%
\bibitem [{\citenamefont {J.~Hsieh}\ and\ \citenamefont
  {Chuu}(2004)}]{Hsieh04}%
  \BibitemOpen
  \bibfield  {author} {\bibinfo {author} {\bibfnamefont {C.~L.}\ \bibnamefont
  {J.~Hsieh}}\ and\ \bibinfo {author} {\bibfnamefont {D.}~\bibnamefont
  {Chuu}},\ }\href@noop {} {\bibfield  {journal} {\bibinfo  {journal} {Chin. J.
  Phys.}\ }\textbf {\bibinfo {volume} {42}},\ \bibinfo {pages} {585} (\bibinfo
  {year} {2004})}\BibitemShut {NoStop}%
\bibitem [{\citenamefont {Shenvi}\ \emph {et~al.}(2003)\citenamefont {Shenvi},
  \citenamefont {Brown},\ and\ \citenamefont {Whaley}}]{Shenvi04}%
  \BibitemOpen
  \bibfield  {author} {\bibinfo {author} {\bibfnamefont {N.}~\bibnamefont
  {Shenvi}}, \bibinfo {author} {\bibfnamefont {K.}~\bibnamefont {Brown}}, \
  and\ \bibinfo {author} {\bibfnamefont {K.}~\bibnamefont {Whaley}},\
  }\href@noop {} {\bibfield  {journal} {\bibinfo  {journal} {Phys. Rev. A}\
  }\textbf {\bibinfo {volume} {68}},\ \bibinfo {pages} {052313} (\bibinfo
  {year} {2003})}\BibitemShut {NoStop}%
\bibitem [{\citenamefont {Regev}\ and\ \citenamefont {Schiff}(2008)}]{Regev08}%
  \BibitemOpen
  \bibfield  {author} {\bibinfo {author} {\bibfnamefont {O.}~\bibnamefont
  {Regev}}\ and\ \bibinfo {author} {\bibfnamefont {L.}~\bibnamefont {Schiff}},\
  }\href@noop {} {\bibfield  {journal} {\bibinfo  {journal} {Lecture Notes in
  Computer Science}\ }\textbf {\bibinfo {volume} {5125}},\ \bibinfo {pages}
  {773} (\bibinfo {year} {2008})}\BibitemShut {NoStop}%
\bibitem [{\citenamefont {Salas}(2008)}]{Salas08}%
  \BibitemOpen
  \bibfield  {author} {\bibinfo {author} {\bibfnamefont {P.~J.}\ \bibnamefont
  {Salas}},\ }\href@noop {} {\bibfield  {journal} {\bibinfo  {journal} {Eur.
  Phys. J. D}\ }\textbf {\bibinfo {volume} {46}},\ \bibinfo {pages} {365}
  (\bibinfo {year} {2008})}\BibitemShut {NoStop}%
\bibitem [{\citenamefont {Vrana}\ \emph {et~al.}(2014)\citenamefont {Vrana},
  \citenamefont {Reeb}, \citenamefont {Reitzner},\ and\ \citenamefont
  {Wolf}}]{Vrana13}%
  \BibitemOpen
  \bibfield  {author} {\bibinfo {author} {\bibfnamefont {P.}~\bibnamefont
  {Vrana}}, \bibinfo {author} {\bibfnamefont {D.}~\bibnamefont {Reeb}},
  \bibinfo {author} {\bibfnamefont {D.}~\bibnamefont {Reitzner}}, \ and\
  \bibinfo {author} {\bibfnamefont {M.~M.}\ \bibnamefont {Wolf}},\ }\href@noop
  {} {\bibfield  {journal} {\bibinfo  {journal} {New J. Phys.}\ }\textbf
  {\bibinfo {volume} {16}},\ \bibinfo {pages} {073033} (\bibinfo {year}
  {2014})}\BibitemShut {NoStop}%
\bibitem [{\citenamefont {Temme}(2014)}]{Temme14}%
  \BibitemOpen
  \bibfield  {author} {\bibinfo {author} {\bibfnamefont {K.}~\bibnamefont
  {Temme}},\ }\href@noop {} {\bibfield  {journal} {\bibinfo  {journal}
  {arXiv:1404.1977[quant-ph]}\ } (\bibinfo {year} {2014})}\BibitemShut
  {NoStop}%
\bibitem [{\citenamefont {Datta}\ and\ \citenamefont {Shaji}(2012)}]{Datta12}%
  \BibitemOpen
  \bibfield  {author} {\bibinfo {author} {\bibfnamefont {A.}~\bibnamefont
  {Datta}}\ and\ \bibinfo {author} {\bibfnamefont {A.}~\bibnamefont {Shaji}},\
  }\href@noop {} {\bibfield  {journal} {\bibinfo  {journal} {Modern Physics
  Letters B}\ }\textbf {\bibinfo {volume} {26}},\ \bibinfo {pages} {1230010}
  (\bibinfo {year} {2012})}\BibitemShut {NoStop}%
\bibitem [{\citenamefont {Farhi}\ and\ \citenamefont
  {Gutmann}(1998)}]{Farhi98}%
  \BibitemOpen
  \bibfield  {author} {\bibinfo {author} {\bibfnamefont {E.}~\bibnamefont
  {Farhi}}\ and\ \bibinfo {author} {\bibfnamefont {S.}~\bibnamefont
  {Gutmann}},\ }\href@noop {} {\bibfield  {journal} {\bibinfo  {journal} {Phys.
  Rev. A}\ }\textbf {\bibinfo {volume} {57}},\ \bibinfo {pages} {2403}
  (\bibinfo {year} {1998})}\BibitemShut {NoStop}%
\bibitem [{\citenamefont {Zalka}(1999)}]{Zalka99}%
  \BibitemOpen
  \bibfield  {author} {\bibinfo {author} {\bibfnamefont {C.}~\bibnamefont
  {Zalka}},\ }\href@noop {} {\bibfield  {journal} {\bibinfo  {journal} {Phys.
  Rev. A}\ }\textbf {\bibinfo {volume} {60}},\ \bibinfo {pages} {2746}
  (\bibinfo {year} {1999})}\BibitemShut {NoStop}%
\bibitem [{\citenamefont {Fujiwara}\ and\ \citenamefont
  {Imai}(2008)}]{Fujiwara2008}%
  \BibitemOpen
  \bibfield  {author} {\bibinfo {author} {\bibfnamefont {A.}~\bibnamefont
  {Fujiwara}}\ and\ \bibinfo {author} {\bibfnamefont {H.}~\bibnamefont
  {Imai}},\ }\href {\doibase 10.1088/1751-8113/41/25/255304} {\bibfield
  {journal} {\bibinfo  {journal} {J. Phys. A: Math. Theor.}\ }\textbf {\bibinfo
  {volume} {41}},\ \bibinfo {pages} {255304} (\bibinfo {year}
  {2008})}\BibitemShut {NoStop}%
\bibitem [{\citenamefont {Demkowicz-Dobrzanski}\ and\ \citenamefont
  {Maccone}(2014)}]{Demkowicz14}%
  \BibitemOpen
  \bibfield  {author} {\bibinfo {author} {\bibfnamefont {R.}~\bibnamefont
  {Demkowicz-Dobrzanski}}\ and\ \bibinfo {author} {\bibfnamefont
  {L.}~\bibnamefont {Maccone}},\ }\href@noop {} {\bibfield  {journal} {\bibinfo
   {journal} {arXiv:1407.2934 [quant-ph]}\ } (\bibinfo {year}
  {2014})}\BibitemShut {NoStop}%
\bibitem [{\citenamefont {Ulam-Orgikh}\ and\ \citenamefont
  {Kitagawa}(2001)}]{Orgikh2001}%
  \BibitemOpen
  \bibfield  {author} {\bibinfo {author} {\bibfnamefont {D.}~\bibnamefont
  {Ulam-Orgikh}}\ and\ \bibinfo {author} {\bibfnamefont {M.}~\bibnamefont
  {Kitagawa}},\ }\href {\doibase 10.1103/PhysRevA.64.052106} {\bibfield
  {journal} {\bibinfo  {journal} {Phys. Rev. A}\ }\textbf {\bibinfo {volume}
  {64}},\ \bibinfo {pages} {052106} (\bibinfo {year} {2001})}\BibitemShut
  {NoStop}%
\bibitem [{\citenamefont {Bures}(1969)}]{Bures69}%
  \BibitemOpen
  \bibfield  {author} {\bibinfo {author} {\bibfnamefont {D.}~\bibnamefont
  {Bures}},\ }\href@noop {} {\bibfield  {journal} {\bibinfo  {journal} {Trans.
  Am. Math. Soc.}\ }\textbf {\bibinfo {volume} {135}},\ \bibinfo {pages} {199}
  (\bibinfo {year} {1969})}\BibitemShut {NoStop}%
\bibitem [{\citenamefont {Helstrom}(1967)}]{Helstrom67}%
  \BibitemOpen
  \bibfield  {author} {\bibinfo {author} {\bibfnamefont {C.~W.}\ \bibnamefont
  {Helstrom}},\ }\href@noop {} {\bibfield  {journal} {\bibinfo  {journal}
  {Phys. Lett. A}\ }\textbf {\bibinfo {volume} {25}},\ \bibinfo {pages} {101}
  (\bibinfo {year} {1967})}\BibitemShut {NoStop}%
\bibitem [{\citenamefont {Bengtsson}\ and\ \citenamefont
  {Zyczkowski}(2006)}]{Bengtsson2006}%
  \BibitemOpen
  \bibfield  {author} {\bibinfo {author} {\bibfnamefont {I.}~\bibnamefont
  {Bengtsson}}\ and\ \bibinfo {author} {\bibfnamefont {K.}~\bibnamefont
  {Zyczkowski}},\ }\href@noop {} {\emph {\bibinfo {title} {Geometry of quantum
  states: an introduction to quantum entanglement}}}\ (\bibinfo  {publisher}
  {Cambridge Univeristy Press},\ \bibinfo {year} {2006})\BibitemShut {NoStop}%
\bibitem [{\citenamefont {Braunstein}\ and\ \citenamefont
  {Caves}(1994)}]{Braunstein94}%
  \BibitemOpen
  \bibfield  {author} {\bibinfo {author} {\bibfnamefont {S.~L.}\ \bibnamefont
  {Braunstein}}\ and\ \bibinfo {author} {\bibfnamefont {C.~M.}\ \bibnamefont
  {Caves}},\ }\href@noop {} {\bibfield  {journal} {\bibinfo  {journal} {Phys.
  Rev. Lett.}\ }\textbf {\bibinfo {volume} {72}},\ \bibinfo {pages} {3439}
  (\bibinfo {year} {1994})}\BibitemShut {NoStop}%
\bibitem [{\citenamefont {Giovannetti}\ \emph {et~al.}(2003)\citenamefont
  {Giovannetti}, \citenamefont {Lloyd},\ and\ \citenamefont
  {Maccone}}]{Giovannetti03}%
  \BibitemOpen
  \bibfield  {author} {\bibinfo {author} {\bibfnamefont {V.}~\bibnamefont
  {Giovannetti}}, \bibinfo {author} {\bibfnamefont {S.}~\bibnamefont {Lloyd}},
  \ and\ \bibinfo {author} {\bibfnamefont {L.}~\bibnamefont {Maccone}},\
  }\href@noop {} {\bibfield  {journal} {\bibinfo  {journal} {Phys. Rev. A}\
  }\textbf {\bibinfo {volume} {67}},\ \bibinfo {pages} {052109} (\bibinfo
  {year} {2003})}\BibitemShut {NoStop}%
\bibitem [{\citenamefont {Taddei}\ \emph {et~al.}(2013)\citenamefont {Taddei},
  \citenamefont {Escher}, \citenamefont {Davidovich},\ and\ \citenamefont
  {de~Matos~Filho}}]{Taddei13}%
  \BibitemOpen
  \bibfield  {author} {\bibinfo {author} {\bibfnamefont {M.~M.}\ \bibnamefont
  {Taddei}}, \bibinfo {author} {\bibfnamefont {B.~M.}\ \bibnamefont {Escher}},
  \bibinfo {author} {\bibfnamefont {L.}~\bibnamefont {Davidovich}}, \ and\
  \bibinfo {author} {\bibfnamefont {R.~L.}\ \bibnamefont {de~Matos~Filho}},\
  }\href@noop {} {\bibfield  {journal} {\bibinfo  {journal} {Phys Rev. Lett.}\
  }\textbf {\bibinfo {volume} {110}},\ \bibinfo {pages} {050402} (\bibinfo
  {year} {2013})}\BibitemShut {NoStop}%
\bibitem [{\citenamefont {T\'oth}\ and\ \citenamefont {Petz}(2013)}]{Toth2013}%
  \BibitemOpen
  \bibfield  {author} {\bibinfo {author} {\bibfnamefont {G.}~\bibnamefont
  {T\'oth}}\ and\ \bibinfo {author} {\bibfnamefont {D.}~\bibnamefont {Petz}},\
  }\href {\doibase 10.1103/PhysRevA.87.032324} {\bibfield  {journal} {\bibinfo
  {journal} {Phys. Rev. A}\ }\textbf {\bibinfo {volume} {87}},\ \bibinfo
  {pages} {032324} (\bibinfo {year} {2013})}\BibitemShut {NoStop}%
\bibitem [{\citenamefont {Yu}(2013)}]{Yu2013}%
  \BibitemOpen
  \bibfield  {author} {\bibinfo {author} {\bibfnamefont {S.}~\bibnamefont
  {Yu}},\ }\href@noop {} {\bibfield  {journal} {\bibinfo  {journal} {ArXiv
  e-prints}\ } (\bibinfo {year} {2013})},\ \Eprint
  {http://arxiv.org/abs/1302.5311} {arXiv:1302.5311 [quant-ph]} \BibitemShut
  {NoStop}%
\bibitem [{\citenamefont {Meyer}\ and\ \citenamefont {Wong}(2014)}]{Meyer14}%
  \BibitemOpen
  \bibfield  {author} {\bibinfo {author} {\bibfnamefont {D.~A.}\ \bibnamefont
  {Meyer}}\ and\ \bibinfo {author} {\bibfnamefont {T.~G.}\ \bibnamefont
  {Wong}},\ }\href@noop {} {\bibfield  {journal} {\bibinfo  {journal} {Phys.
  Rev. A}\ }\textbf {\bibinfo {volume} {89}},\ \bibinfo {pages} {012312}
  (\bibinfo {year} {2014})}\BibitemShut {NoStop}%
\end{thebibliography}%

\end{document}